\documentclass[11pt]{article}
\usepackage{a4wide,cite,graphicx,latexsym,amssymb}
\usepackage{color}

\newcommand{\rd}{{\rm d}}
\newcommand{\wt}{\widetilde}

\newcommand{\RE}{\mbox{\rm Re}}
\newcommand{\eqn}[1]{(\ref{#1})}
\newcommand{\mev}{\mbox{\rm MeV}}
\newcommand{\gev}{\mbox{\rm GeV}}

\begin{document}


\begin{titlepage}

\phantom{x}\vspace{-1cm}

\begin{flushright}
{UAB-FT-642}\\
{IFIC/08-16}\\
{FTUV/08-0312}\\[16mm]
\end{flushright}

\begin{center}
{\Large\sf\bf What can be learned from the Belle\\[2mm]
 spectrum for the decay \boldmath{$\tau^-\to\nu_\tau K_S\pi^-$}}\\[12mm]

{\large\bf Matthias Jamin${}^{1}$, Antonio Pich${}^{2}$, and}
{\large\bf Jorge Portol\'es${}^{2}$}\\[10mm]

{\small\sl ${}^{1}$ Instituci\'o Catalana de Recerca i Estudis
 Avan\c{c}ats (ICREA)} and\\
{\small\sl Departament de F\'{\i}sica Te\`orica, IFAE, UAB, E-08193 Bellaterra,
 Barcelona, Spain}\\[3mm]
{\small\sl ${}^{2}$ Departament de F\'{\i}sica Te\`orica, IFIC,
 Universitat de Val\`encia -- CSIC,}\\
{\small\sl Apartat de Correus 22085, E-46071 Val\`encia, Spain.}\\[2cm] 
\end{center}

{\bf Abstract:}
A theoretical description of the differential decay spectrum for the decay
$\tau^-\to\nu_\tau K_S\pi^-$, which is based on the contributing $K\pi$
vector and scalar form factors $F_+^{K\pi}(s)$ and $F_0^{K\pi}(s)$ being
calculated in the framework of resonance chiral theory (R$\chi$T), additionally
imposing constraints from dispersion relations as well as short distance QCD,
provides a good representation of a recent measurement of the spectrum by the
Belle collaboration. Our fit allows to deduce the total branching fraction
$B[\tau^-\to\nu_\tau K_S\pi^-]=0.427 \pm 0.024\,\%$ by integrating the spectrum,
as well as the $K^*$ resonance parameters $M_{K^*}=895.3 \pm 0.2\,$MeV and
$\Gamma_{K^*}=47.5 \pm 0.4\,$MeV, where the last two errors are statistical
only. From our fits, we confirm that the scalar form factor $F_0^{K\pi}(s)$
is required to provide a good description, but we were unable to further
constrain this contribution. Finally, from our results for the vector form
factor $F_+^{K\pi}(s)$, we update the corresponding slope and curvature
parameters $\lambda_+^{'}=(25.2\pm 0.3)\cdot 10^{-3}$ and
$\lambda_+^{''}=(12.9\pm 0.3)\cdot 10^{-4}$, respectively.

\vfill

\noindent
PACS: 13.35.Dx, 11.30.Rd, 11.55.Fv

\noindent
Keywords: Decays of taus, chiral symmetries, dispersion relations
\end{titlepage}

\newpage
\setcounter{page}{1}


\section{Introduction}

An ideal system to study low-energy QCD under rather clean conditions is
provided by hadronic decays of the $\tau$ lepton
\cite{bnp92,bra89,bra88,np88,pich89}. Detailed investigations of the $\tau$
hadronic width as well as invariant mass distributions allow to determine a
plethora of QCD parameters, a most prominent example being the QCD coupling
$\alpha_s$. Furthermore, the experimental separation of the Cabibbo-allowed
decays and Cabibbo-suppressed modes into strange particles
\cite{dhz05,opal04,aleph99} opened a means to also determine the quark-mixing
matrix element $|V_{us}|$ \cite{gjpps07,gjpps04,gjpps03} as well as the mass
of the strange quark \cite{bck04,cdghpp01,dchpp01,km00,kkp00,pp99,ckp98,pp98},
additional fundamental parameters within the Standard Model, from the $\tau$
strange spectral function.

The dominant contribution to the Cabibbo-suppressed $\tau$ decay rate arises
from the decay $\tau\to\nu_\tau K\pi$. The corresponding distribution function
has been measured experimentally in the past by ALEPH \cite{aleph99} and OPAL
\cite{opal04}. More recently, high-statistics data for the
$\tau\to\nu_\tau K\pi$ spectrum became available from the Belle experiment
\cite{belle07}, and results for the total branching fraction are also available
from BaBar \cite{babar07,ban07}, with good prospects for results on the
spectrum from BaBar and BESIII in the near future.

These new results call for a refined theoretical understanding of the
$\tau\to\nu_\tau K\pi$ decay spectrum, and in ref.~\cite{jpp06} we have
provided a description based on the chiral theory with resonances (R$\chi$T)
\cite{egpr89,eglpr89}, under the additional inclusion of constraints from
dispersion relations. To start with, the general expression for the
differential decay distribution takes the form \cite{fm96}
\begin{equation}
\label{dGtau2kpi}
\frac{\rd\Gamma_{K\pi}}{\rd\sqrt{s}} \,=\, \frac{G_F^2|V_{us}|^2 M_\tau^3}
{32\pi^3s}\,S_{\rm EW}\biggl(1-\frac{s}{M_\tau^2}\biggr)^2\Biggl[ \biggl(1+2\,
\frac{s}{M_\tau^2}\biggr) q_{K\pi}^3\,|F_+^{K\pi}(s)|^2 +
\frac{3\Delta_{K\pi}^2}{4s}\,q_{K\pi}|F_0^{K\pi}(s)|^2 \Biggr] ,
\end{equation}
where we have assumed isospin invariance and have summed over the two possible
decays $\tau^-\to\nu_\tau\overline K^0\pi^-$ and $\tau^-\to\nu_\tau K^-\pi^0$,
with the individual decay channels contributing in the ratio
$2\hspace{-0.4mm}:\!1$ respectively.  In this expression, $S_{\rm EW}$ is an
electro-weak correction factor, $F_+^{K\pi}(s)$ and $F_0^{K\pi}(s)$ are the
vector and scalar $K\pi$ form factors respectively which will be explicated in
more detail in section~2. Furthermore, $\Delta_{K\pi}\equiv M_K^2-M_\pi^2$,
and $q_{K\pi}$ is the kaon momentum in the rest frame of the hadronic system,
\begin{equation}
q_{K\pi}(s) \,=\, \frac{1}{2\sqrt{s}}\sqrt{\Big(s-(M_K+M_\pi)^2\Big)
\Big(s-(M_K-M_\pi)^2\Big)}\cdot\theta\Big(s-(M_K+M_\pi)^2\Big) \,.
\end{equation}

By far the dominant contribution to the decay distribution originates from the
$K^*(892)$ meson. In the next section, we shall recall the effective description
of this contribution to the vector form factor $F_+^{K\pi}(s)$ in the framework
of R$\chi$T that was presented in ref.~\cite{jpp06}, quite analogous to a
similar description of the pion form factor given in
refs.~\cite{gp97,pp01,scp02}. A second vector resonance, namely the $K^*(1410)$
meson, can straightforwardly be included in the effective chiral description.
Finally, the scalar $K\pi$ form factor $F_0^{K\pi}(s)$ was calculated in the
same R$\chi$T plus dispersive constraint framework in a series of articles
\cite{jop00,jop01a,jop01b}, and the recent update of $F_0^{K\pi}(s)$
\cite{jop06} will be incorporated in our work as well.

Based on the theoretical expression \eqn{dGtau2kpi} for the spectrum and the
form factors discussed in section~2, in section~3, we shall perform fits of
our description to the Belle data \cite{belle07} for the decay
$\tau^-\to\nu_\tau K_S\pi^-$. From these fits it follows that both the scalar
contribution and the second vector resonance are required in order to obtain
a good description of the experimental spectrum. In addition, the fits allow
to determine the resonance parameters of the charged $K^*(892)$ and $K^*(1410)$
mesons. Finally, integrating the distribution function
$\rd\Gamma_{K\pi}/\rd\sqrt{s}$, we are also in a position to present results
for the total $B[\tau^-\to\nu_\tau K_S\pi^-]$ branching fraction.

\section{The form factors}

A theoretical representation of the vector form factor $F_+^{K\pi}(s)$, which
is based on fundamental principles, has been developed in ref.~\cite{jpp06}, 
in complete analogy to the description of the pion form factor presented in
refs.~\cite{gp97,pp01,scp02}. This approach employed our present knowledge on
effective hadronic theories, short-distance QCD, the large-$N_C$ expansion as
well as analyticity and unitarity. For the pion form factor the resulting
expressions provide a very good description of the experimental data
\cite{gp97,pp01,scp02}.

Precisely following the approach of ref.~\cite{gp97}, in \cite{jpp06} we found
the following representation of the form factor $F_+^{K\pi}(s)$:
\begin{equation}
\label{FpEFT}
F_+^{K\pi}(s) \,=\, \frac{M_{K^*}^2 {\rm e}^{{\frac{3}{2}\RE\big[
\widetilde{H}_{K\pi}(s)+\widetilde{H}_{K\eta}(s)\big]}}}{M_{K^*}^2 -
s - iM_{K^*}\Gamma_{K^*}(s)} \;.
\end{equation}
The one-loop function $\widetilde{H}(s)$ is related to the corresponding
function $H(s)$ of \cite{gl85b} by $\widetilde{H}(s)\equiv H(s) -
2L_9^r\,s/(3F_0^2)\approx[s M^r(s)-L(s)]/(F_K F_\pi)$.\footnote{In our
expressions, we have decided to replace all factors of $1/F_0^2$ by
$1/(F_K F_\pi)$ since for the $K\pi$ system it is to be expected that
higher-order chiral corrections lead to the corresponding renormalisation
of the meson decay constant.} Explicit expressions for $M^r(s)$ and $L(s)$
can be found in ref.~\cite{gl85}.  The one-loop function $\widetilde{H}(s)$
depends on the chiral scale $\mu$, and in eq.~\eqn{FpEFT}, this scale should
be taken as $\mu=M_{K^*}$. In ref.  \cite{GomezDumm:2000fz}, the off-shell
width of a vector resonance was defined through the two-point vector current
correlator, performing a Dyson-Schwinger resummation within R$\chi$T
\cite{egpr89,eglpr89}.  Following this scheme the energy-dependent width
$\Gamma_{K^*}(s)$ is found to be
\begin{equation}
\label{GaKs}
\Gamma_{K^*}(s) \,=\, \frac{G_V^2 M_{K^*} s}{64\pi F_K^2 F_\pi^2} \Big[
\sigma_{K\pi}^3(s) + \sigma_{K\eta}^3(s) \Big] \,=\,
\Gamma_{K^*}\frac{s}{M_{K^*}^2}\,\frac{\left[\sigma_{K\pi}^3(s) +
\sigma_{K\eta}^3(s) \right]}{\left[\sigma_{K\pi}^3(M_{K^*}^2) +
\sigma_{K\eta}^3(M_{K^*}^2) \right]} \,,
\end{equation}
where $\Gamma_{K^*}\equiv \Gamma_{K^*}(M_{K^*}^2)$, and $G_V$ is the chiral
vector coupling which appears in the framework of the R$\chi$T \cite{egpr89}.
The phase space function $\sigma_{K\pi}(s)$ is given by
$\sigma_{K\pi}(s)=2q_{K\pi}(s)/\sqrt{s}$, and $\sigma_{K\eta}(s)$ follows
analogously with the replacement $M_\pi\to M_\eta$. Re-expanding eq.~\eqn{FpEFT}
in $s$ and comparing to the corresponding $\chi$PT expression \cite{gl85b}, in
the SU(3) symmetry limit one reproduces the short-distance constraint for the
vector coupling $G_V=F_0/\sqrt{2}$ \cite{eglpr89} which guarantees a vanishing
form factor at $s$ to infinity, as well as the lowest-resonance estimate.

Since the $\tau$ lepton can also decay hadronically into the second vector
resonance $K^{*'}\equiv K^*(1410)$, this particle has been included in our
parametrisation of the vector form factor $F_+^{K\pi}(s)$. A parametrisation
which is motivated by the R$\chi$T framework \cite{egpr89,eglpr89} can be
written as follows:
\begin{equation}
\label{FpKsKsp}
F_+^{K\pi}(s) \,=\, \Biggl[\, \frac{M_{K^*}^2+\gamma\,s}{M_{K^*}^2 - s -
iM_{K^*}\Gamma_{K^*}(s)} - \frac{\gamma\,s}{M_{K^{*'}}^2 - s -
iM_{K^{*'}}\Gamma_{K^{*'}}(s)} \,\Biggr] {\rm e}^{{\frac{3}{2}
\RE\big[\widetilde{H}_{K\pi}(s)+\widetilde{H}_{K\eta}(s)\big]}}  \,.
\end{equation}
This parametrisation incorporates all known constraints from $\chi$PT and
R$\chi$T. At low energies, it reproduces eq.~\eqn{FpEFT} up to corrections
proportional to $\gamma\,s\,(M_{K^*}-M_{K^{*'}})$. The relation of the
parameter $\gamma$ to the R$\chi$T couplings takes the form
$\gamma = F_V G_V/(F_K F_\pi) - 1$, when one assumes a vanishing form factor
at large $s$ in the $N_C$ to infinity limit. It is difficult, to a priori
asses a precise value for $\gamma$, but below we shall be able to fit it from
the comparison of our description with the Belle spectrum. The width of the
second resonance cannot be set unambiguously. Therefore, we have decided to
endow the $K^*(1410)$ contribution with a generic width as expected for a
vector resonance. Hence, $\Gamma_{K^{*'}}(s)$ will be taken to have the form
\begin{equation}
\Gamma_{K^{*'}}(s) \,=\, \Gamma_{K^{*'}}\,\frac{s}{M_{K^{*'}}^2}\,
\frac{\sigma_{K\pi}^3(s)}{\sigma_{K\pi}^3(M_{K^{*'}}^2)} \,.
\end{equation}

As a final ingredient for a prediction of the differential decay distribution
of the decay $\tau\to\nu_\tau K\pi$ according to eq.~\eqn{dGtau2kpi}, we
require the scalar form factor $F_0^{K\pi}(s)$. This form factor was calculated
in a series of articles \cite{jop00,jop01a,jop01b} in the framework of
R$\chi$T, again also employing constraints from dispersion theory as well as
the short-distance behaviour.\footnote{The original motivation for a precise
description of $F_0^{K\pi}(s)$ was the determination of the strange quark mass
$m_s$ from scalar sum rules, also performed in \cite{jop01b}.} Quite recently,
the determination of $F_0^{K\pi}(s)$ was updated in \cite{jop06} by employing
novel experimental constraints on the form factor at the Callan-Treiman point
$\Delta_{K\pi}$, and in our fits below, we shall also make use of this update.

A remaining question is which value to use for the form factors $F_+^{K\pi}(s)$
and $F_0^{K\pi}(s)$ at the origin. However, inspecting eq.~\eqn{dGtau2kpi}, one
realises that what is needed is not $F_+^{K\pi}(0)=F_0^{K\pi}(0)$ itself, but
only the product $|V_{us}| F_+^{K\pi}(0)$. Once this normalisation is fixed,
in the fits we only need to determine the shape of reduced form factors
$\wt F_+^{K\pi}(s)$ and $\wt F_0^{K\pi}(s)$ which are normalised to one at
the origin:
\begin{equation}
\wt F_+^{K\pi}(s) \,\equiv\, \frac{F_+^{K\pi}(s)}{F_+^{K\pi}(0)} \,, \qquad
\wt F_0^{K\pi}(s) \,\equiv\, \frac{F_0^{K\pi}(s)}{F_+^{K\pi}(0)} \,.
\end{equation}
This also entails, that after fixing the normalisation of the decay spectrum
by giving a value to $|V_{us}| F_+^{K\pi}(0)$, we are in a position to predict
the total branching fraction $B[\tau^-\to\nu_\tau K_S\pi^-]$ just from a fit
of the shape of the form factors, independent of normalisation issues.

The product $|V_{us}| F_+^{K\pi}(0)$ is determined most precisely from the
analysis of semi-leptonic kaon decays. The most recent average was presented
by the FLAVIAnet kaon working group, and reads \cite{FKWG08}
\begin{equation}
\label{VusF0}
|V_{us}| F_+^{K^0\pi^-}(0) \,=\, 0.21664 \pm 0.00048 \,.
\end{equation}
In what follows, we have renormalised our description for the form factors to
one and have assumed the result \eqn{VusF0} for the global normalisation.
Incidentally, the value in \eqn{VusF0} already corresponds to the $K^0\pi^-$
channel which was analysed by the Belle collaboration \cite{belle07}.
Therefore, possible isospin-breaking corrections to the normalisation are
already properly taken into account.

\section{Fits to the Belle \boldmath{$\tau\to\nu_\tau K\pi$} spectrum}

For our fits to the decay spectrum of the $\tau^-\to\nu_\tau K_S\pi^-$
transition as obtained by the Belle collaboration \cite{belle07}, we make
the following Ansatz:
\begin{equation}
\label{FitFun}
\frac{1}{2}\cdot \frac{2}{3}\cdot 0.0115\,[{\rm GeV/bin}]\,{\cal N}_T\cdot
\frac{1}{\Gamma_\tau \bar B_{K\pi}}\,\frac{\rd\Gamma_{K\pi}}{\rd\sqrt{s}} \,.
\end{equation}
The factors $1/2$ and $2/3$ come from the fact that the $K_S\pi^-$
channel has been analysed. Then, $11.5\,$MeV was the bin-width chosen by the
Belle collaboration, and ${\cal N}_T=53110$ the total number of observed
signal events. Finally, $\Gamma_\tau$ is the total decay width of the $\tau$
lepton and $\bar B_{K\pi}$ a remaining normalisation factor that will be
deduced from the fits. The normalisation of our Ansatz \eqn{FitFun} is
taken such that for a perfect agreement between data and fit function,
$\bar B_{K\pi}$ would correspond to the total branching fraction
$B_{K\pi}\equiv B[\tau^-\to\nu_\tau K_S\pi^-]$ which is obtained by integrating
the decay spectrum. Differences between $\bar B_{K\pi}$ and $B_{K\pi}$ point
to imperfections of the fit, and will constitute one source of systematic
uncertainties. As we shall see further below, for better fits also the
agreement between $\bar B_{K\pi}$ and $B_{K\pi}$ improves as expected.

Before entering the details of our fits, let us discuss the numerical values
of all input parameters. For the meson masses, we employ the physical masses
corresponding to the decay channel in question, namely $M_{K_S}=497.65\,$MeV,
$M_{\pi^-}=139.57\,$MeV and $M_\eta=547.51\,$MeV \cite{pdg06}. For the meson
decay constants, we use the findings of the recent review \cite{rs08}, in our
normalisation that is $F_\pi=92.3\,$MeV and $F_K/F_\pi=1.196$. For the
electro-weak correction factor, we have utilised the result for inclusive
hadronic $\tau$ decays, $S_{\rm EW}=1.0201$ \cite{gjpps04} (and references
therein). Even though the electro-weak correction factor for the exclusive
decay in question need not be the same as $S_{\rm EW}$, to the precision we
are working this choice is supposedly sufficient. Besides, we are not aware
of a published result for the correct factor in the case of the exclusive
decay studied here. All remaining input parameters which have not been
mentioned explicitly, are taken according to their PDG values \cite{pdg06}.

As an initial step, only the central $K^*$ resonance region is fitted, in order
to get an idea about the $K^*$ resonance parameters. For this fit, two forms
of the dominant vector form factor $F_+^{K\pi}(s)$ are used. On the one hand,
we employ our description \eqn{FpEFT} as discussed in the last section. On the
other hand, we also investigate a pure Breit-Wigner resonance shape as was used
in the experimental work of the Belle collaboration. This later allows a better
comparison to the findings of ref.~\cite{belle07}. The Breit-Wigner resonance
factor is defined by
\begin{equation}
\label{BWs}
BW_{K^*}(s) \,\equiv\, \frac{M_{K^*}^2}{M_{K^*}^2-s-i M_{K^*}
\Gamma_{K^*}(s)} \,,
\end{equation}
where the energy dependent width $\Gamma_{K^*}(s)$ takes the form
\begin{equation}
\label{Gas}
\Gamma_{K^*}(s) \,=\, \Gamma_{K^*}\,\frac{s}{M_{K^*}^2}\,
\frac{\sigma_{K\pi}^3(s)}{\sigma_{K\pi}^3(M_{K^*}^2)} \,.
\end{equation}
Thus, the $K^*$ width of \eqn{Gas} coincides with eq.~\eqn{GaKs} if the $K\eta$
contribution is neglected. Although our equations \eqn{BWs} and \eqn{Gas} are
written in a form different from the one employed in \cite{belle07}, the
expressions are in agreement. The Breit-Wigner version of the $K\pi$ vector
form factor $F_+^{K\pi}(s)$ then reads
\begin{equation}
F_+^{K\pi}(s) \,=\, F_+^{K\pi}(0)\,BW_{K^*}(s) \,.
\end{equation}
In practice, as discussed above, for our fits we only require the reduced form
factor $\wt F_+^{K\pi}(s)$ which in this case is equal to the Breit-Wigner
factor $BW_{K^*}(s)$.

\begin{figure}[thb]
\begin{center}
\includegraphics[angle=0, width=15cm]{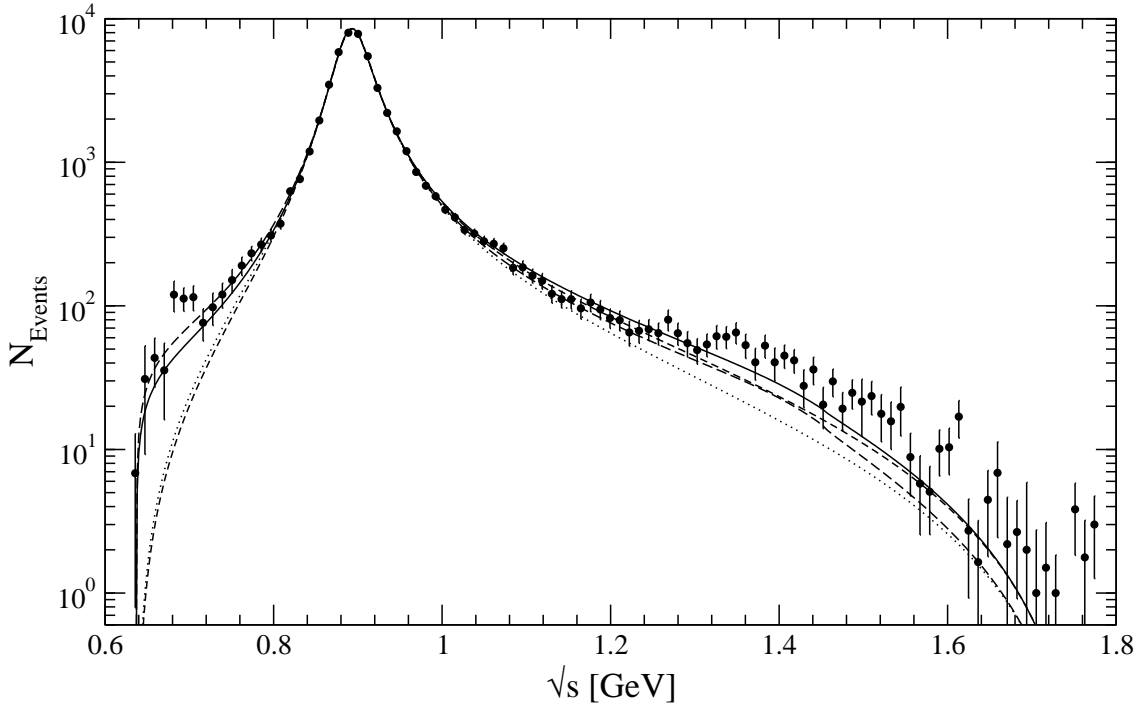}
\end{center}
\caption{Fit result for the differential decay distribution of the decay
$\tau\to\nu_\tau K\pi$, when fitted with a pure $K^*$ vector resonance (dotted
and short-dashed curves) or with $K^*$ plus the central scalar form factor
$F_0^{K\pi}(s)$ as given in \cite{jop06} (long-dashed and solid curves).
\label{fig1}}
\end{figure}

For our first fit, we employ the Belle data \cite{belle07} in the range
$0.808\,$--$\,1.015\,$GeV (data points $16\,$--$\,34$), where the vector form
factor dominates and should provide a good description. The resulting fit
parameters are presented as the left-hand column in table~\ref{tab1} for the
Breit-Wigner fit, and the right hand column for the chiral fit. Graphically,
the corresponding fits are shown as the dotted and short-dashed lines in
figure~\ref{fig1} respectively, together with the experimental data points.
The fitted $K^*$ mass $M_{K^*}$ for the Breit-Wigner fit is close to the result
by the Belle collaboration \cite{belle07}, while the width $\Gamma_{K^*}$ is
found to be somewhat larger. Besides the normalisation factor $\bar B_{K\pi}$,
in table~\ref{tab1} we have also listed in brackets the result for the branching
fraction $B_{K\pi}$ that would be obtained when integrating the spectrum. The
$\chi^2/$n.d.f. for this fit is found to be of order 2. Nevertheless, later we
shall see that our final fit including all contributions will have a
$\chi^2/$n.d.f. of order 1. So this is nothing to worry about at this point.
From figure~\ref{fig1}, one observes that the fit provides a reasonable
description of the data in the fit region, but both, much below and much above
the resonance peak marked deviations are clearly visible, implying missing
contributions that will be discussed below.

\begin{table}[bht]
\renewcommand{\arraystretch}{1.1}
\begin{center}
\begin{tabular}{ccc}
\hline
 & BW form for $F_+^{K\pi}(s)$ & Chiral form for $F_+^{K\pi}(s)$ \\
\hline
$\bar B_{K\pi}\;(B_{K\pi})$ &
$0.3435 \pm 0.0042\,\%\;(0.3311\,\%)$ & $0.4658 \pm 0.0057\,\%\;(0.4541\,\%)$\\
$M_{K^*}$ & $895.59 \pm 0.18\;\mev$ & $894.93 \pm 0.18\;\mev$ \\
$\Gamma_{K^*}$ & $48.06 \pm 0.45\;\mev$ & $47.47 \pm 0.44\;\mev$ \\
\hline
$\chi^2/$n.d.f. & 30.3/16 & 30.8/16 \\
\hline
\end{tabular}
\end{center}
\caption{Fit to the Belle $\tau\to\nu_\tau K\pi$ spectrum in the $K^*$
resonance region with a pure vector resonance shape.\label{tab1}}
\end{table}

Performing in an analogous fashion the fit to the Belle data with the R$\chi$T
form of $F_+^{K\pi}(s)$, the obtained fit parameters are listed in the
right-hand column in table~\ref{tab1}, and the fit curve is displayed as the
short-dashed line in figure~\ref{fig1}. The parameters obtained from both fits
differ to some extent, especially the normalisation $\bar B_{K\pi}$, due to the
different functional forms of the vector form factor. Still, we will postpone
a detailed discussion of our numerical results until presenting the complete
fit including all contributions below.\footnote{As the fit is practically
insensitive to the parameter $r$ in the Blatt-Weisskopf barrier factor
appearing in our previous parametrisation of the $K^*$ width \cite{jpp06}, we
have decided to set $r$ to zero, so that our fits are more directly comparable
to the fits performed by the Belle collaboration \cite{belle07}, who have not
applied such a factor. Employing the central result of our previous fit
$r = 3.5\;\gev^{-1}$ \cite{jpp06} would give practically the same $\chi^2$,
but would result in a $K^*$ mass that is about $1.4\;\mev$ lower and a $K^*$
width about $0.8\;\mev$ lower. These conclusions are the same for both the
Breit-Wigner or chiral form of the vector form factor $F_+^{K\pi}(s)$. This
observation again indicates the fact that the precise functional form of the
vector form factor matters for the resulting values of $K^*$ mass $M_{K^*}$
and width $\Gamma_{K^*}$.} From figure~\ref{fig1}, we see that while both,
the chiral and the Breit-Wigner fits give a similar spectrum below the $K^*$
resonance peak, above the peak there are substantial differences. This will
play an important role below, when we shall aim at improving the fit by adding
a second vector resonance $K^{*'}$, because it will certainly influence its
fit parameters.

\begin{table}[htb]
\renewcommand{\arraystretch}{1.1}
\begin{center}
\begin{tabular}{ccc}
\hline
 & BW form for $F_+^{K\pi}(s)$ & Chiral form for $F_+^{K\pi}(s)$ \\
\hline
$\bar B_{K\pi}\;(B_{K\pi})$ &
$0.3575 \pm 0.0041\,\%\;(0.3518\,\%)$ & $ 0.4767 \pm 0.0056\,\%\;(0.4726\,\%)$\\
$M_{K^*}$ & $895.56 \pm 0.18\;\mev$ & $894.92 \pm 0.18\;\mev$ \\
$\Gamma_{K^*}$ & $47.05 \pm 0.42\;\mev$ & $46.94 \pm 0.42\;\mev$ \\
\hline
$\chi^2/$n.d.f. & 43.5/28 & 46.2/28 \\
\hline
\end{tabular}
\end{center}
\caption{Fit to the Belle $\tau\to\nu_\tau K\pi$ spectrum in the $K^*$
resonance region with a vector resonance shape for $F_+^{K\pi}(s)$ and the
central scalar form factor $F_0^{K\pi}(s)$.\label{tab2}}
\end{table}

Thus far, we have completely omitted the contribution of the scalar $K\pi$
form factor $F_0^{K\pi}(s)$ to the differential $\tau\to\nu_\tau K\pi$ decay
spectrum. When adding the corresponding contribution with the central
parameters as presented in \cite{jop06}, it is found that the combined
theoretical spectrum gives a good description also in the region below the
$K^*$ resonance, with the exception of three data points in the range
$0.682\,$--$\,0.705\,$GeV (points $5,6,7$). Therefore, as our next fit, we fit
the entire low-energy region $0.636\,$--$\,1.015\,$GeV while keeping the
scalar form factor $F_0^{K\pi}(s)$ fixed but leaving out in the fit the
problematic data points $5,6$ and $7$.\footnote{These three data points look
as if there might be a bumpy structure, perhaps related to the $K_0^*(800)$.
However, as has been discussed in section~7 of \cite{jop00}, the $K_0^*(800)$
is in fact present in our chiral description of the scalar form factor. As it
is rather broad, we see no way how one could accommodate such a bump in the
low-energy region below the $K^*$ resonance.} The resulting fit parameters
in the case of the Breit-Wigner and chirally inspired vector form factor
$F_+^{K\pi}(s)$ are tabulated in table \ref{tab2}, and the corresponding fit
curves are plotted as the long-dashed and solid lines in figure~\ref{fig1}
respectively. From table~\ref{tab2} one observes that $M_{K^*}$ is almost
unchanged, the width $\Gamma_{K^*}$ is slightly decreased, and also the
$\chi^2/$n.d.f. is somewhat reduced, although it is still larger than roughly
$1.5$. Nevertheless, it is clear that the scalar contribution is required in
order to give a more satisfactory description of the region below the $K^*$
resonance.

\begin{table}[htb]
\renewcommand{\arraystretch}{1.1}
\begin{center}
\begin{tabular}{ccc}
\hline
 & BW form for $F_+^{K\pi}(s)$ & Chiral form for $F_+^{K\pi}(s)$ \\
\hline
$\bar B_{K\pi}\;(B_{K\pi})$ &
$0.423 \pm 0.012\,\%\;(0.421\,\%)$ & $0.430 \pm 0.011\,\%\;(0.427\,\%)$ \\
$M_{K^*}$ & $895.12 \pm 0.19\;\mev$ & $895.28 \pm 0.20\;\mev$ \\
$\Gamma_{K^*}$ & $46.79 \pm 0.41\;\mev$ & $47.50 \pm 0.41\;\mev$ \\
$M_{K^{*'}}$ & $1598 \pm 25\;\mev$ & $1307 \pm 17\;\mev$ \\
$\Gamma_{K^{*'}}$ & $224 \pm 47\;\mev$ & $206 \pm 49\;\mev$ \\
$\beta$, $\gamma$ & $-\,0.079 \pm 0.010$ & $-\,0.043 \pm 0.010$ \\
\hline
$\chi^2/$n.d.f. & 88.7/81 & 79.5/81 \\
\hline
\end{tabular}
\end{center}
\caption{Full fit to the Belle $\tau\to\nu_\tau K\pi$ spectrum with the two
$K^*$ and $K^{*'}$ vector resonances in $F_+^{K\pi}(s)$ and the central scalar
form factor $F_0^{K\pi}(s)$.\label{tab3}}
\end{table}

\begin{figure}[thb]
\begin{center}
\includegraphics[angle=0, width=15cm]{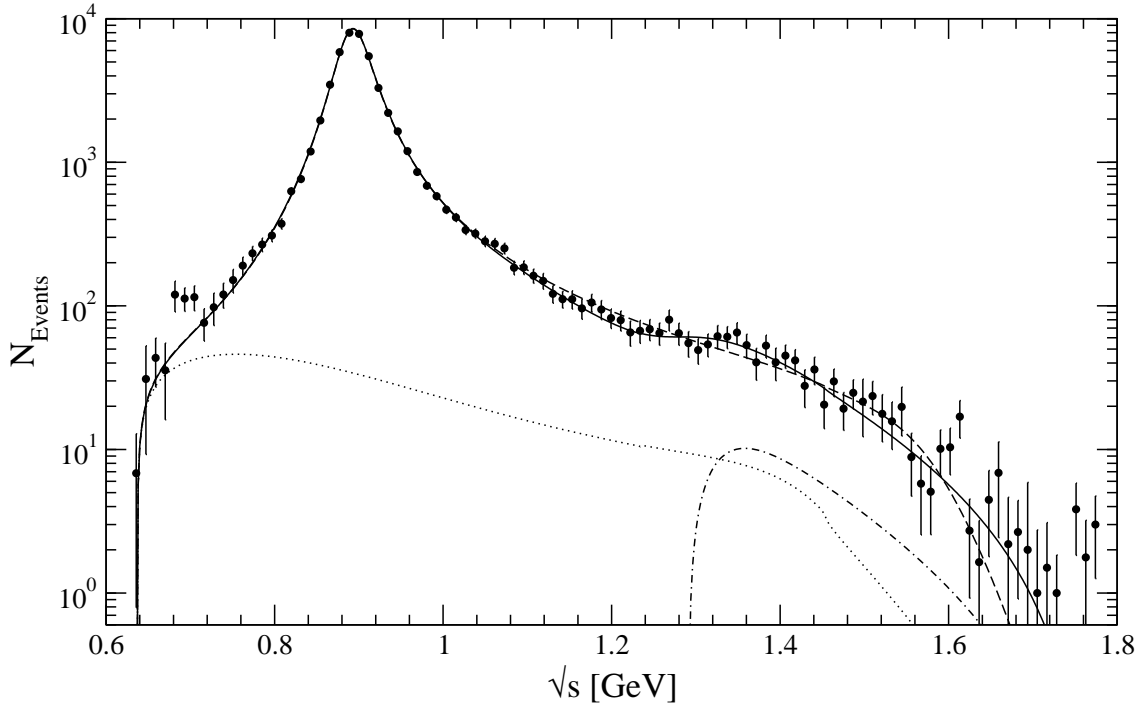}
\end{center}
\caption{Main fit result to the Belle data \cite{belle07} for the differential
decay distribution of the decay $\tau^-\to\nu_\tau K_S\pi^-$. Our theoretical
description includes the Breit-Wigner (dashed line) or R$\chi$T (solid line)
vector form factors with two resonances, as well as the scalar form factor
according to ref.~\cite{jop06}. For R$\chi$T also the scalar (dotted line) and
$K^{*'}$ (dashed-dotted) contributions are displayed.\label{fig2}}
\end{figure}

As the last step, now we also improve upon the description of the region
above the $K^*$ resonance by including as a second vector resonance the
$K^{*'}$. In the case of the Breit-Wigner form factor, the inclusion of the
$K^{*'}$ resonance can be achieved by writing
\begin{equation}
\label{fp2bw}
F_+^{K\pi}(s) \,=\, \frac{F_+^{K\pi}(0)}{1+\beta}\,\Big[\, BW_{K^*}(s) +
\beta\,BW_{K^{*'}}(s) \,\Big] \,,
\end{equation}
whereas in the case of the chiral resonance description, the corresponding
expression for $F_+^{K\pi}(s)$ including the $K^{*'}$ is given above in
eq.~\eqn{FpKsKsp} and depends on the mixing parameter $\gamma$. Again, our
fits are displayed in a graphical form in figure~\ref{fig2}, where the solid
line corresponds to the R$\chi$T description, and the dashed line to the fit
with a vector form factor according to eq.~\eqn{fp2bw}. For R$\chi$T, in
addition we have separately displayed the contributions of the scalar form
factor (dotted line) and of the $K^{*'}$ resonance (dashed-dotted line). The
resulting fit parameters have been collected in table~\ref{tab3}. We observe
that the chirally inspired description of ref.~\cite{jpp06} provides the better
fit, and that as expected the $K^{*'}$ mass $M_{K^{*'}}$ turns out to be very
different, though the $\Gamma_{K^{*'}}$ widths (probably by chance) agree
rather well. The mixing parameters $\beta$ and $\gamma$ also differ, but due
to the different functional forms of our two descriptions for the vector form
factor $F_+^{K\pi}(s)$, anyway they cannot be compared.

Up to now, in our fits we have only employed the central prediction for the
scalar form factor $F_0^{K\pi}(s)$. Thus the question arises what happens if
we modify $F_0^{K\pi}(s)$. As the normalisation of the form factors can be
fixed by experiment, we only require the shape of $F_0^{K\pi}(s)$ and for this,
in our dispersive approach \cite{jop06,jop01a,jop01b}, the dominant input
parameter is the value of the ratio $F_0^{K\pi}(\Delta_{K\pi})/F_0^{K\pi}(0)$
at the Callan-Treiman point $\Delta_{K\pi}\equiv M_K^2-M_\pi^2$, which has been
discussed in detail in \cite{jop06}. We can then introduce a fit parameter
$\alpha$ which describes the change of shape of $F_0^{K\pi}(s)$ when
$F_0^{K\pi}(\Delta_{K\pi})/F_0^{K\pi}(0)$ is modified. Let $\alpha=0$
correspond to our central result of \cite{jop06}, $\alpha=1$ to the scalar
form factor which arises when $F_0^{K\pi}(\Delta_{K\pi})/F_0^{K\pi}(0)$ is
larger by $1\sigma$, and $\alpha=-1$ when
$F_0^{K\pi}(\Delta_{K\pi})/F_0^{K\pi}(0)$ is smaller by $1\sigma$. Adding
$\alpha$ to our fit parameters, for the chirally inspired $F_+^{K\pi}(s)$
we obtain $\alpha = 4.4 \pm 1.9$, and for the pure Breit-Wigner form
$\alpha = 6.3 \pm 2.7$, with only a slight change of the other parameters
and a small improvement in the $\chi^2/$n.d.f. From this we conclude that
the fit prefers a slightly larger $F_0^{K\pi}(s)$, but the sensitivity to
$\alpha$ is not very strong. Furthermore, the largest changes when leaving
$\alpha$ free are in the parameters of the $K^{*'}$, which entails that the
found values for $\alpha$ are driven by the energy region above the $K^*$
resonance, where the theoretical description is less well founded. If the
same exercise is repeated with the fits which only include the low-energy
and $K^*$ resonance region (fits of table~\ref{tab2}), then we obtain
$\alpha = 4.7 \pm 7.9$ in the case of the R$\chi$T description. Hence, with
the present precision of the data and in particular the open question about
the three data points in the low-energy region, we are not able to further
constrain the contribution of the scalar form factor $F_0^{K\pi}(s)$.

Let us now come to a detailed discussion of our central fit results of
table~\ref{tab3}. The $\chi^2/$n.d.f. of both the chiral and the Breit-Wigner
fits is of the order of one, but nevertheless the chiral fit provides the
better description of the experimental data. For the complete fit including
two vector resonances and the scalar contribution, within the fit uncertainties
the normalisation $\bar B_{K\pi}$ and the branching fraction $B_{K\pi}$ are in
very good agreement. In addition, as can be observed from table~\ref{tab3},
also the branching fractions extracted from the two versions of parametrising
$F_+^{K\pi}(s)$ display perfect consistency, once all contributions have been
included in the fit. The remaining small difference can be traced back to
the exponential factor in the numerator of the R$\chi$T expression \eqn{FpEFT}.
Since our chiral model for $F_+^{K\pi}(s)$ is theoretically better motivated
and furthermore provides the better fit quality, as our central result for
the branching fraction, we quote:
\begin{equation}
\label{BKpi}
B[\tau^-\to\nu_\tau K_S\pi^-] \,=\, 0.427 \pm 0.011 \pm 0.021 \,\%
\,=\, 0.427 \pm 0.024 \,\% \,.
\end{equation}
The first quoted error corresponds to the statistical fit uncertainty. To be
conservative, in the second error we made an attempt to estimate systematic
uncertainties. To this end, we have performed analogous fits, where the
chiral factors $1/F_0^2$ are taken to be $1/F_\pi^2$, which should give an
idea of the importance of higher-order chiral corrections. (See footnote~1.)
Then the branching fraction for the full R$\chi$T fit turns out to be
$B_{K\pi}=0.448\,\%$, and we take the difference of this result to our main
value as an additional systematic uncertainty. When comparing to previous
determinations, within the uncertainties our result \eqn{BKpi} is in agreement
with the findings of the Belle collaboration
$B[\tau^-\to\nu_\tau K_S\pi^-] = 0.404\,\pm\,0.013 \,\%$ \cite{belle07},
which are just based on a pure counting of events, as well as the Particle
Data Group average for the related branching fraction
$B[\tau^-\to\nu_\tau \bar K^0\pi^-] = 0.90\,\pm\,0.04 \,\%$ \cite{pdg06}.
When assuming isospin invariance, the above results can also be compared with
the BaBar measurement
$B[\tau^-\to\nu_\tau K^-\pi^0] = 0.416\,\pm\,0.018 \,\%$ \cite{babar07},
showing very good overall consistency.

As far as the parameters of the charged $K^*$ resonance are concerned, within
the uncertainties our value $M_{K^*}=895.3\pm0.2$ is in very good agreement
with the Belle result \cite{belle07}. However, it is about $3.5\,$MeV larger
than the current PDG average \cite{pdg06}.\footnote{Funnily enough, it is in
much better agreement with the PDG average for the {\em neutral} $K^*$ mass.
For more details, the reader is referred to the related discussion of the $K^*$
mass in ref.~\cite{belle07}.} On the other hand, our finding for the width
$\Gamma_{K^*}=47.5\pm0.4\,$MeV is significantly lower than the PDG average,
but still roughly $1\,$MeV larger than the Belle result. The corresponding
value of the chiral coupling $G_V$ which appears in eq.~\eqn{GaKs} is found
to be $G_V=72.0\pm0.6\,$MeV. For the second vector resonance, the $K^*(1410)$,
the obtained mass from our central chiral fit is about $100\,$MeV lower than
the PDG average \cite{pdg06}, while for the width, we find reasonable agreement
to the PDG value. However, for the Breit-Wigner fit $M_{K^{*'}}$ was found
to turn out much larger, which implies that the mass of the $K^*(1410)$
strongly depends on our parametrisation of the form factors and its
determination is therefore not very reliable. As a general remark, we like to
emphasise that one should not compare or average determinations done with
different functional parametrisations.

\section{Conclusions}

Let us briefly summarise our findings before drawing further conclusions.
From a description of the $K\pi$ vector and scalar form factors $F_+^{K\pi}(s)$
and $F_0^{K\pi}(s)$ in the framework R$\chi$T, additionally imposing constraints
from dispersion relations as well as short distance QCD, we were able to obtain
a good fit to the recent Belle data \cite{belle07} for the spectrum of the decay
$\tau^-\to\nu_\tau K_S\pi^-$. From our fit we could extract the corresponding
branching fraction and the parameters of the $K^*$ resonance
\begin{equation}
B[\tau^-\to\nu_\tau K_S\pi^-] \,=\, 0.427 \pm 0.024 \,\% \,,
\end{equation}
\begin{equation}
M_{K^*} \,=\, 895.3 \pm 0.2 \;\mev \,, \quad
\Gamma_{K^*} \,=\, 47.5 \pm 0.4 \;\mev \,,
\end{equation}
where the quoted errors for $M_{K^*}$ and $\Gamma_{K^*}$ only include the
statistical fit uncertainties. Besides, we observe a substantial model
dependence of these parameters. (See footnote~3.) This model dependence is
even more pronounced for the second included resonance, the $K^*(1410)$, and
therefore we are unable to make a reliable prediction for $M_{K^{*'}}$ and
$\Gamma_{K^{*'}}$.

As far as the scalar form factor $F_0^{K\pi}(s)$ is concerned, below the
$K^*$ resonance it is obvious that this contribution is required in order to
provide a satisfactory description of the data. (With the exception of three
data points which appear to form a small bump.) Trying to also fit the scalar
part, it is seen that the data prefer a slightly larger contribution, but on
the basis of the present data this is statistically not significant. Above
the $K^*$, we have the well established $K_0^*(1430)$ resonance, but here it
interferes with higher vector resonances. Due to these correlations and the
strong model dependence of the higher vector resonances, it will be difficult
to disentangle scalar and vector contributions without a dedicated analysis
of angular correlations \cite{km92,fm96}.

An independent investigation of the Belle $\tau^-\to\nu_\tau K_S\pi^-$ decay
spectrum on the basis of Mushkelishvili-Omn\'es integral equations, also
incorporating chiral constraints at low energies as well as QCD short-distance
constraints at high energies was recently published in ref.~\cite{mou07}.
A visual inspection of the corresponding fit results presented in figure~5 of
\cite{mou07} suggests that the quality of the fit is not as good as in our case,
though no further details, e.g.~a~$\chi^2$, were provided in \cite{mou07}.
Still, it would be interesting to see, if somehow the approaches used in
ref.~\cite{mou07} and in our work could be merged, to be able to impose as
many theoretical constraints as possible on the employed form factors.

Already in ref.~\cite{jpp06}, from our description of the vector form factor
$F_+^{K\pi}(s)$, we deduced slope and curvature of the form factor close to
$s=0$, which are important parameters in the determination of $|V_{us}|$ from
$K_{l3}$ decays. Let us define a general expansion of the reduced form factor
$\wt F_+^{K\pi}(s)$ as:
\begin{equation}
\wt F_+^{K\pi}(s) \,\equiv\, 1 + \lambda_+^{'}\frac{s}{M_{\pi^-}^2} +
\frac{1}{2}\, \lambda_+^{''}\frac{s^2}{M_{\pi^-}^4} +
\frac{1}{6}\,\lambda_+^{'''} \frac{s^3}{M_{\pi^-}^6} + \ldots \,,
\end{equation}
where $\lambda_+^{'}$, $\lambda_+^{''}$ and $\lambda_+^{'''}$ are the slope,
curvature and cubic expansion parameter respectively. On the basis of our fit
results of table~\ref{tab3}, we are now in a position to update these
quantities, also estimating the corresponding uncertainties, which yields:
\begin{equation}
\label{lambdap}
\lambda_+^{'}   \,=\, (25.20 \pm 0.33)\cdot 10^{-3} \,, \;\;
\lambda_+^{''}  \,=\, (12.85 \pm 0.31)\cdot 10^{-4} \,, \;\;
\lambda_+^{'''} \,=\,  (9.56 \pm 0.28)\cdot 10^{-5} \,.
\end{equation}
In an attempt to estimate systematic uncertainties from higher orders in the
chiral expansion, like in the last section we have again also investigated
the case $F_K=F_\pi$, which contributes the largest part of the error quoted
in \eqn{lambdap}. The next important source of uncertainty stems from the
mixing parameter of the $K^{*'}$ resonance $\gamma$, for which we have used
the fit result of table~\ref{tab3}. Besides, the vector masses $M_{K^*}$ and
$M_{K^{*'}}$ have been varied by $1\,$MeV and $100\,$MeV respectively, but
these modifications only have a small impact on the uncertainties for the
expansion parameters of $\wt F_+^{K\pi}(s)$. Comparing to the most recent
determination of $\lambda_+^{'}$ and $\lambda_+^{''}$ from an average of
current experimental results for $K_{l3}$ decays \cite{FKWG08} (where also
detailed references to the individual experiments can be found), we observe
that both determinations are in very good agreement, though for the time
being our theoretical extraction is more precise.

To conclude, our R$\chi$T description of the $K\pi$ vector and scalar form
factors provides a good representation of the experimental data of the Belle
collaboration for the spectrum of the decay $\tau^-\to\nu_\tau K_S\pi^-$
\cite{belle07}, thereby allowing to deduce many parameters of this approach.
The used method can also be applied to $\tau$ decay channels which involve
three final state hadrons, and this has already been performed successfully
for the decays $\tau\to\nu_\tau\pi\pi\pi$ \cite{gpp03} as well as
$\tau\to\nu_\tau K K\pi$ \cite{por07}. In the near future, we plan to return
to the still missing decay mode $\tau\to\nu_\tau K\pi\pi$, which is the most
interesting one in view of getting a better handle on the hadronic $\tau$
decay rate into strange final states.

\vskip 1cm \noindent
{\Large\bf Acknowledgements}

\noindent
We are most grateful to the Belle collaboration, in particular S.~Eidelman,
D.~Epifanov and B.~Shwartz, for providing their data and for useful discussions.
This work has been supported in part by EU Contract No. MRTN-CT-2006-035482
(FLAVIAnet) (MJ, AP, JP), by CICYT-FEDER-FPA2005-02211 (MJ), by MEC (Spain)
grant No. FPA2007-60323 (AP, JP) and by Spanish Consolider-Ingenio 2010
Programme CPAN (CSD2007-00042).


\end{document}